\documentclass[showpacs,preprintnumbers,amsmath,amssymb,      
twocolumn, tightenlines,prl
]{revtex4}

\usepackage[dvips]{graphicx}
\usepackage{bm}
\usepackage{amsmath}
\usepackage{dcolumn}

\begin{document}

\bibliographystyle{apsrev}

\title{Reflection from black holes}

    \author{M.Yu.Kuchiev}
    \email[Email:]{kmy@newt.phys.unsw.edu.au}
    \affiliation{ School of Physics, University of New South Wales,
      Sydney 2052, Australia}
    

\begin{abstract} 
  Black holes are presumed to have an ideal ability to absorb and keep
  matter. Whatever comes close to the event horizon, a boundary
  separating the inside region of a black hole from the outside world,
  inevitably goes in and remains inside forever. This work shows,
  however, that quantum corrections make possible a surprising
  process, reflection: a particle can bounce back from the event
  horizon. For low energy particles this process is efficient, black
  holes behave not as holes, but as mirrors, which changes our
  perception of their physical nature. Possible ways for observations
  of the reflection and its relation to the Hawking radiation process
  are outlined.
\end{abstract}

    \pacs {04.70.Dy, 04.20.Gz}

    \maketitle
    
    Conventional intuitive arguments attribute two defining properties
    to black holes. They should absorb every particle that comes
    close, releasing nothing back. There is, however, a limitation on
    this intuitive picture. It stems from thermodynamics properties
    that ascribe the entropy and temperature to black holes. The first
    indication that gravitational fields could have entropy came when
    investigation by Christodoulou \cite{christodoulou_1970} of the
    Penrose process \cite{penrose_1969} for extracting energy from the
    Kerr black hole showed that there is a quantity that could not go
    down, which Hawking found \cite{hawking_1970} to be proportional
    to the area of the horizon. Further research of Bardeen {\it et
      al} demonstrated \cite{bardeen_carter_hawking_1970} that black
    holes should obey laws similar to the laws of thermodynamics. An
    important step made by Bekenstein \cite{%
      bekenstein_1972,bekenstein_1973,bekenstein_1974} indicated that
    the area was actually the physical entropy. This observation was
    supported and enriched by the discovery of the Hawking radiation
    \cite{hawking_1974,hawking_1975} and Unruh process
    \cite{unruh_1976}. For a recent review on the thermodynamics
    approach to black holes see \cite{wald_2001}, see also
    \cite{frolov_novikov_1998,thorne_1994,chandrasekhar_1993} for a
    comprehensive discussion of black hole properties. The black hole
    temperature and the corresponding radiation process limit an
    ability of black holes to keep matter. Let us show that their
    ability for absorption is also limited, they are capable of
    reflection. This fact changes our perception of their physical
    nature.

\begin{figure}
\centering
 \includegraphics[height=8cm,keepaspectratio=true]{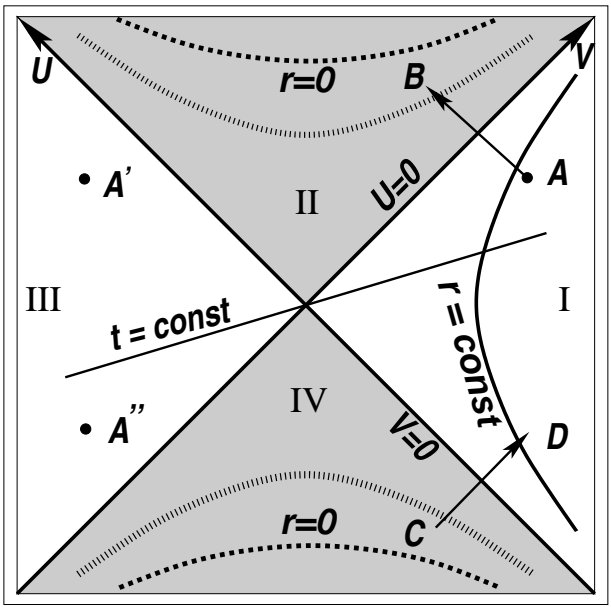}
\caption{Kruskal coordinates. Areas {\it I\/} and {\it III\/} represent
  two identical copies of the outside region; {\it II\/}, {\it IV\/}
  show two inside regions. Hyperbolic curves $UV = const$ describe
  condition $r=const$, the dotted curve shows location of $r=0$, the
  inclined straight line presents condition $t=const$.  The direction
  of time arrow in {\it I\/} and {\it III\/} is opposite. The incoming
  particle follows {\it AB\/} crossing the horizon $U=0$ and residing
  in {\it II\/}. The outgoing particle {\it CD\/} escapes from {\it
    IV\/} crossing the horizon $V=0$ and coming to {\it I\/}.  Areas
  {\it II\/} and {\it IV\/} are not connected, which ensures classical
  confinement in {\it II\/}.  The symmetrically located points {\it
    A,A',A''} are used to reveal the symmetry (\ref{T2p}) of the
  space-time.  The wave function (\ref{gen}) describes mixing of
  events corresponding to incoming and outgoing trajectories
  ({\it AB\/} and {\it CD\/}), which results in phenomena of
  reflection and radiation.  }\label{one}
\end{figure}

      Consider the simplest case of the black hole described by the
      Schwarzschild geometry
    \begin{equation}
      \label{schw}
      ds^2\!
      = \!-\left(1\!-\! \frac{1}{r}\right)dt^2 + \frac{dr^2}{1-1/r}
    + r^2 (d \theta^2 + \sin^2 \! \theta \,d\varphi^2)\,.
      \end{equation}
      The relativistic units $\hbar=c=1$, supplemented by condition
      $2Gm =1$ are use, where $G$ and $m$ are the gravitational
      constant and the mass of the black hole.  The event horizon for
      the geometry (\ref{schw}) is located on the sphere with the
      radius $r = r_g \equiv 1$; the
      condition $r > 1$ ($r < 1$) specifies the outside (inside)
      region of the black hole. Consider a probing particle with the
      mass $\mu$, describing its classical motion in the gravitational
      field by the Hamilton-Jacobi classical equations,
      $g^{\kappa\lambda} \partial _\kappa {\cal S} \partial _\lambda
      {\cal S} = -\mu^2$, where ${ \cal S}$ is the classical action.
      The metric defined in (\ref{schw}) allows full separation of
      variables, ${\cal S} = -\varepsilon t + L \phi + S(r)$, where
      $\varepsilon$, $L$ and $\phi$ are the energy, orbital momentum
      and azimuthal angle of the particle. The radial part of the
      action $S(r)$ satisfies equation derived from (\ref{schw})
      \begin{eqnarray}
        \label{S'}
 S'(r) \!=\!\mp
\left[ \varepsilon^2 \!-\!  \left( \mu^2\! + \! \frac{L^2}{r^2} \right)
\left(1\!-\! \frac{1}{r} \right) \right]^{1/2}
\!\! \frac{1}{  1\!-\! 1/r }\,.
      \end{eqnarray}
      It follows from this that the action has a logarithmic
      singularity on the horizon
      \begin{eqnarray}
        \label{ln}
S(r) \simeq \mp \,
 \varepsilon \,\ln( r -1 ),  \quad   r \rightarrow 1~.
      \end{eqnarray}
      The signs minus and plus in (\ref{S'}) and (\ref{ln}) correspond
      to the incoming and outgoing trajectories respectively; they are
      discussed in some detail below with the help of Fig.\ref{one}.
      Importantly, the singularity of the action is a covariant
      property, it persists even in those coordinates that eliminate
      the singularity of the metric and classical equations of
      motion.  Consider, for instance, the Kruskal \cite{kruskal_1960}
      coordinates $U,V$ defined according to
      \begin{eqnarray}
        \label{U}
U = \!&-& \!(r -1)^{1/2} \exp[\,(r - t)/2\,]~,\\ 
\label{V}
V =\! &\,\,&  \!(r -1)^{1/2} \exp [\,(r + t)/2\,]~.
      \end{eqnarray}
      It is known, see e.g. Ref.\cite{misner_thorne_wheeler_1973},
      that in $U,V$ coordinates there are four regions of the
      space-time that are shown in Fig. \ref{one}. Areas {\it I\/} and
      {\it III\/} represent two identical copies of the outside,
      asymptotically flat region.  Areas {\it II\/}, {\it IV\/} show
      two copies of the inside region that have different physical
      properties.  All four areas are divided by the horizon that is
      located at $U=0$, or $V=0$.  In the vicinity of the horizon the
      incoming particle follows the trajectory {\it AB\/} and, after
      crossing the horizon, resides in {\it II\/}.  There are also the
      outgoing trajectories; following the trajectory {\it CD\/} the
      particle escapes from the inside region {\it IV\/} into the
      outside world {\it I\/}.  Importantly, the areas {\it II\/} and
      {\it IV\/} are not connected. This fact ensures that when the
      particle comes to the inside region {\it II} it stays there,
      there is no classical trajectory that would bring it back into
      the outside world.
     \noindent 
     In Kruscal coordinates the metric is regular on the horizon.  The
     classical equations of motion are also regular. In the vicinity
     of the horizon they are described by simple equations $V=const$
     for the incoming trajectory, and $U=const$ for the outgoing
     trajectory (Fig.\ref{one}).  In contrast, the singularity of the
     action persists.  It is convenient to present it in terms of the
     full action $\mathcal{S}=-\varepsilon t + S(r)$ (in which we drop
     now the irrelevant for us angular term $L\varphi$), which in the
     vicinity of the horizon reads
      \begin{eqnarray}
        \label{ah1}
\mathcal{S} \simeq &-& \varepsilon \ln(V^2), ~\quad U^2 \ll
      V^2 \ll  1~, \\ \label{ah2}
\mathcal{S} \simeq &\,\,&  \varepsilon \ln(V^2), ~\quad V^2 \ll
        U^2 \ll 1~.
      \end{eqnarray}
      We observe an interesting distinction. The classical equations
      of motion can be made regular on the horizon, while the action
      has a singularity. Similar difference exists in the Aharonov-Bohm
      effect \cite{aharonov_bohm_1959}, where the classical
      equations of motion do not depend on the flux of the magnetic
      field, while the action exhibits the singularity. Using the
      Aharonov-Bohm effect as a motivation, one can anticipate that the
      singularity of the action should have important consequences for
      quantum description. Developing this argument, let us describe the
      radial motion of the probing particle with the help of the
      semiclassical wave function $\phi(r) \propto \exp[ i S(r)]$
      \footnote {It can be shown that on the horizon $r \rightarrow 1
        $ the semiclassical description is asymptotically accurate; in
        particular, the preexponential factor in relation $\phi(r)
        \propto \exp[ i S(r)]$ is a constant.} Taking $r$ in the
      outside region in the vicinity of the horizon and using the
      action (\ref{ln}) one can write the wave function as
      \begin{eqnarray}
        \label{R}
\phi (r) = &\exp[-i\,\varepsilon
      \ln (\,r -1\,)\, ]& +  \\ \nonumber 
\mathcal{R}& \exp [\,\,\, 
i \,\varepsilon \ln( \,r -1\,) \,]&.        
      \end{eqnarray}
      Here the two terms on the right-hand side are constructed from
      the incoming and outgoing classical trajectories respectively.
      Hence, the first term represents the proper incoming wave that
      describes the particle approaching the event horizon. The
      corresponding classical trajectory is shown by {\it AB} in
      Fig.\ref{one}.  The second term presents the proper outgoing
      wave; it is introduced in order to allow for an opportunity of
      the mixing, interference of the incoming and outgoing waves in
      the wave function.  This wave corresponds to the classical
      trajectory shown by {\it CD} in Fig.\ref{one}.  The quantity
      $\mathcal{R} $ in (\ref{R}) has the meaning of the reflection
      coefficient. The conservation of the flux of particles implies
      $|R| \le 1$. Moreover, conventional intuitive arguments prompt
      one to put the reflection coefficient in (\ref{R}) to zero,
      $R=0$; indeed, if the black hole is a perfect absorber, then
      only the incoming wave should describe the particle approaching
      the horizon.  However, the wave function (\ref{R}) is singular
      on the horizon $r = 1$, while the intuitive arguments are based
      on the properties of classical trajectories that by themselves
      are unable to probe a possible singularity at $r=1$, simply
      because the trajectories possess none.
      
      Thus, generally speaking, the reflection coefficient
      $\mathcal{R}$ in (\ref{R}) may have a nonzero value due to
      quantum reasons. To verify that this really happens, let us use
      the analytical continuation of the wave function. Introduce with
      this purpose a new parameter $z,~z =r-1$, considering $z$ as a
      complex variable. Choose some coordinate in the vicinity of the
      horizon taking $z > 0,~ z \ll 1$.  Rotate now $z$ in the complex
      $z$-plain clockwise over the angle $2\pi$ (anticlockwise
      rotation is not justified, see below). This complex
      transformation brings $z$ back to its original physical value $z
      > 0$. However, since the function (\ref{R}) is singular at $z =
      0$, it acquires some new value, let us call it $\phi^{(2\pi)}
      (r)$.  From (\ref{R}) one derives
\begin{eqnarray}
  \label{2pi}
\phi^{((2\pi))}  (r) = \varrho &\exp& [  -i\, \varepsilon \ln( \,r
        -1\,) \,] + 
\\ \nonumber
( \mathcal{R} /  \varrho ) &\exp& [ \,\,\, i\,\varepsilon \ln( \,r
-1\,) \,]~, 
        \end{eqnarray}
        where $\varrho = \exp(-2 \pi  \varepsilon )$.  The
        analytically continued function $\phi^{(2\pi )} (r)$ satisfies
        the same equation as the initial function $\phi(r)$.
        Moreover, one has to expect the wave function $\phi^{(2\pi )}
        (r)$ to satisfy the same normalization conditions as the
        initial wave function $\phi(r)$. This requires that one of the
        coefficients in (\ref{2pi}), either $\varrho$, or
        $\mathcal{R}/\varrho$ should have an absolute value equal to
        unity, similar to the coefficient unity in front of the first
        term on the right-hand side of (\ref{2pi}).  Since $\varrho <
        1$, one finds that $|\mathcal{R}|/\varrho = 1$, concluding
        that 
        \begin{eqnarray}
          \label{R=}
|\mathcal{R}| = \exp(-2 \pi \varepsilon\,)~.  
        \end{eqnarray}
        This result indicates that the event horizon can reflect
        particles.
        
        The symmetry of the space-time provides a more general and
        reliable way leading to the same conclusion.  Remember, the
        areas {\it I\/} and {\it III\/} in Fig.\ref{one} describe two
        {\it identical} sets of the outside region. This means that a
        transformation that brings an arbitrary point $A$ of the
        region {\it I} into the symmetrically located point $A'$ in
        the region {\it III}, see Fig.\ref{one}, is a symmetry of the
        wave function. The wave function (\ref{R}) is presumed to be a
        scalar, therefore this symmetry can manifest itself only
        through variation of the phase of the wave function. In other
        words, the transformation $A\rightarrow A'$ results in the
        transformation $\phi \rightarrow \phi'=\pm \phi$. Being
        applied twice, this transformation brings the point $A'$ back
        to $A$ and should therefore bring the wave function to its
        initial value. This fact fixes the phase of the transformation
        up to a sign $\pm$.
        
        There is a convenient way to implement this symmetry.  Let us
        use firstly the complex rotation $z\rightarrow \exp(-2\pi
        i)z$.  Eqs.(\ref{U}),(\ref{V}) show that it results in the
        transformation $U\rightarrow -U,~V\rightarrow -V$ that brings
        the point $A$ to $A''$ in Fig.\ref{one}.  After that we can
        transform $A''$ to $A'$ using the operation of the time
        inversion $\hat{\mathrm{T}}$, see (\ref{U}),(\ref{V}). The
        necessity of the time inversion is related to the fact that
        the arrow of time flow in areas {\it I} and {\it III} is
        opposite, see the inclined straight line of constant time in
        Fig.\ref{one}. As ususal, the inversion of time $t\rightarrow
        -t$ in the argument of the wave function should be accompanied
        by the complex conjugation of the function, i.e.  the operator
        of the time inversion is defined as
        $\hat{\mathrm{T}}[\,\phi(r,t)\,]\equiv \phi^*(r,-t)$.
        Combining the $2\pi$-rotation on the complex $z$-plane and the
        time inversion we fulfill the desired transformation of the
        point $A$ to $A'$.  The symmetry of the space-time under this
        transformation implies that
        \begin{eqnarray}
          \label{T2p}
          \hat{\mathrm{T}} [\, \phi^{(2\pi)}\,] = \pm \,\phi~.
        \end{eqnarray}
        For the stationary wave function $\phi \propto
        \exp(-i\varepsilon t)$ this symmetry reads
        \begin{eqnarray}
          \label{sym}
       \left[\, \phi^{(2\pi)} (r)\,\right]^* = \pm \, \phi(r)~.  
        \end{eqnarray}
        Using the wave functions (\ref{R}) and (\ref{2pi}) we find
        that the symmetry (\ref{sym}) requires that $\mathcal{R} = \pm
        \exp(-2\pi \varepsilon )$, in accord with Eq.(\ref{R=}).  An
        alternative derivation of this result
	that relies more heavily on dynamical properties of the
        problem, also supports validity of Eq.(\ref{R=}), as will be
        discussed elswhere. The probability of reflection
        $\mathcal{P}$ satisfies
\begin{eqnarray}
  \label{P}
\mathcal{P} = \mathcal{R}^2 = \exp(-\varepsilon/T \,)~.
       \end{eqnarray}
       Eq.(\ref{R=}) shows that the parameter $T$ that appears in
       (\ref{P}) coincides with the Hawking temperature of the black
       hole
       \begin{eqnarray}
         \label{T}
T = \frac{1}{4\pi} \equiv \frac{1}{4\pi} \,\left(\frac{\hbar c}{ \,r_g}
\right)~. 
       \end{eqnarray}
       In the classical limit $\hbar \rightarrow 0$ the temperature
       $T$ and reflection probability $\mathcal{P}$ vanish, but the
       phenomenon of reflection holds on the quantum level,
       $T,\mathcal{P} > 0$.  Eq.(\ref{P}) was derived
       using analytical properties of the wave function that are close
       to analytical properties of the propagator. The propagator is
       known to be an analytical function in the upper semiplane of
       the complex $U$-plane and lower semiplane of the complex
       $V$-plane \cite{hartle_hawking_1976}. This property makes
       possible the clockwise rotation $z\rightarrow \exp(-2\pi i) z$
       that was used for derivation of Eq.(\ref{2pi}) (while the
       anticlockwise rotation is not be justified). The distance from
       the horizon in our derivation can be made arbitrary small, $|z|
       \rightarrow 0$, indicating that the effect arises due to those
       events that happen strictly on the horizon.
       
       It is useful to rewrite the wave function (\ref{R}) in a more
       general form
       \begin{eqnarray}
\label{gen}
       | \phi \rangle = |\mathrm{in} \rangle  + 
\mathcal{R} |\mathrm{out}\rangle~. 
       \end{eqnarray}
       Here the time-dependent factor $\exp(-i\varepsilon t)$ is
       included in the wave function $ | \phi \rangle =
       \exp(-i\varepsilon t)\phi(r)$. This form allows one to present
       the incoming and outgoing waves in (\ref{gen}) via convenient
       Kruskal coordinates (\ref{U}),(\ref{V})
       \begin{eqnarray}
         \label{in}
         |\mathrm{in} \rangle\,\, & =&
\exp[-i\,  \varepsilon  \ln(V^2) \,]~,
\\ \label{out}
         |\mathrm{out} \rangle  &=&
\exp[\,\,\,\,\,i\,  \varepsilon  \ln(U^2) \,]~.
       \end{eqnarray}
       Our consideration so far was restricted by the outside region.
       Eqs.(\ref{in}),(\ref{out}) provide now an opportunity to verify
       that Eqs.(\ref{gen}),(\ref{R=}) remain valid in the inside
       region as well.  To justify this claim let us use the symmetry
       condition (\ref{T2p}) rewriting in the following form
       \begin{eqnarray}
         \label{t-t}
\left[\,\phi^{(2\pi)} (V,U)\,\right]^* = \pm \,\phi (U,V)~, 
       \end{eqnarray}
       where $\phi (U,V) \equiv \langle U,V| \phi \rangle$ and
       $\phi^{(2\pi)} (U,V)$ is defined with the help of the
       transformation $z \rightarrow \exp (-2\pi i )\,z$ applied to
       the function $\phi (U,V) $ in which the arguments $U,V$ are
       functions of $z$.  Eqs.(\ref{in}),(\ref{out}),(\ref{t-t}) show
       that, indeed, Eqs.(\ref{gen}),(\ref{R=}) for the wave function
       remain valid inside the horizon. (Thus
       Eqs.(\ref{gen}),(\ref{R=}) give a general form for the wave
       function, valid inside and outside the horizon.)
        
       An interesting implication arises for a particle confined in
       the inside region of a black hole.  Classically this particle
       is localized in the region {\it II\/} in Fig.\ref{one}.
       However, quantum result (\ref{gen}) shows that the wave
       function of this particle necessarily incorporates an admixture
       of the second term, which is the outgoing wave.  This wave
       corresponds to outgoing classical trajectories, see {\it CD} in
       Fig.\ref{one}.  In other words, the wave function of the
       confined particle necessarily incorporates an admixture of the
       term that is located in the inside region {\it IV}.  Remember
       that the two internal regions {\it II} and {\it IV} are
       complitely isolated from each other in the classical
       approximation. In the quantum picture they prove to be related
       because events that take place in these two regions interfere
       in wave function (\ref{gen}). Therefore an attempt to localize
       the particle exclusively in the region {\it II } fails; there
       is a finite probability to find this particle in the region
       {\it IV}.  From this region the confined particle can escape
       into the outside world following the classical outgoing
       trajectory, see {\it CD} in Fig.\ref{one}.
       
       We see that confinement inside a black hole cannot be absolute,
       the locked in particle has a chance to run away.  Hence, a
       black hole necessarily creates a flux of particles escaping
       from its inside region. It is natural to identify this flux
       with the Hawking radiation. The validity of this identification
       is supported by the following arguments.  Consider the
       probability of the escape, which equals $W_\mathrm{rad} =
       \mathcal{P} (1-\mathcal{P}) $.  The first factor here
       $\mathcal{P}=\mathcal{R}^2$ describes the probability to
       populate the outgoing state in the wave function (\ref{gen}),
       while the second factor $1-\mathcal{P}$ refers to the
       probability to cross the horizon; it is less than 1 due to
       reflection from the horizon (\ref{P}). Comparing this result
       with the probability of absorption of the incoming particle
       $W_\mathrm{abs}=1-\mathcal{P}$ (where again the reflection is
       responsible for the factor $1-\mathcal{P}$), one finds that the
       ratio of the two probabilities satisfies relation
          \begin{eqnarray}
            \label{WW}
W_\mathrm{rad} /W_\mathrm{abs} = \mathcal{P} =
\exp(-\varepsilon/T\,)~, 
          \end{eqnarray}
          which shows that the black hole remains in equilibrium with
          the field of radiation if the latter has the temperature
          $T$.  Consequently, the black hole radiates as a black body
          with the temperature $T$, in agreement with the Hawking
          effect. Thus Eq.(\ref{gen}) provides a new transparrent way
          to explain the radiation process. The radiation takes place
          simply because particles excape from the inside region of
          black holes.
          
          Quantum corrections described by (\ref{gen}),(\ref{R=})
          amend basic properties of black holes.  Classically an
          incoming particle freely crosses the event horizon going
          inside, but after that cannot go out. Quantum corrections
          make possible the reflection of the incoming particle from
          the horizon.  For low energy particles, $\varepsilon < T$,
          the reflection is strong, a black hole behaves as a mirror.
          For experimental observation of this effect there should be
          found a strong source of low frequency radiation located
          closely to a black hole to make the signal reflected by the
          black hole observable \footnote{To mention one opportunity,
            consider the collapse in a binary system that consists of
            a star and a black hole. The gravitational and
            electromagnetic waves that take the energy out of this
            system have the necessary low frequency spectrum and are
            powerful at latest stages of the collapse. One should
            expect that the reflection of these waves from the black
            hole should manifest itself in intensity as well as in
            spectral and angular distributions of the radiation.}.
          The reflection takes place even for strong gravitational
          fields created by small black holes with the mass comparable
          with the Planck mass. There are known few quantum phenomena
          for strong gravitation fields, including the Hawking
          radiation and a suggestion to quantize the black hole
          spectrum \cite{bekenstein_2002_a,bekenstein_2002_b}; the
          reflection from black holes is a new entrant in the
          strong-field area.  Quantum corrections also make impossible
          complete confinement; a particle locked in inside the
          horizon maintains an opportunity to run away into the
          outside world, providing a new simple explanation for the
          phenomenon of the Hawking radiation. Our discussion was
          restricted by the Schwarzschild metric (\ref{schw}). Very
          similar conclusions hold for charged rotating black holes
	  as will be discussed later.

          Discussions with V. V. Flambaum are appreciated. This work
          was supported by the Australian Research Council.

   \end{document}